\documentclass[journal]{IEEEtran}

\usepackage[utf8]{inputenc}
\usepackage{cite,amsmath,amssymb}

\usepackage{graphicx,epstopdf}
\usepackage{enumitem}
\usepackage{epsfig}
\usepackage{textcomp}
\usepackage{xcolor}
\usepackage{hyperref}
\usepackage{cleveref}
\usepackage{orcidlink}
\usepackage{stfloats}
\usepackage{lipsum}

\allowdisplaybreaks

\IEEEoverridecommandlockouts
\IEEEaftertitletext{\vspace{-1.5\baselineskip}}

\begin{document}

\title{RIS-Assisted Aerial Non-Terrestrial Networks: An Intelligent Synergy with Deep Reinforcement Learning}

\author{%
Muhammad~Umer{~\orcidlink{0009-0001-8751-6100}},
Muhammad~Ahmed~Mohsin{~\orcidlink{0009-0005-2766-0345}},
Aryan~Kaushik{~\orcidlink{0000-0001-6252-4641}},
Qurrat-Ul-Ain~Nadeem{~\orcidlink{0000-0001-8423-3482}},
Ali~Arshad~Nasir{~\orcidlink{0000-0001-5012-1562
}},~and~Syed~Ali~Hassan{~\orcidlink{0000-0002-8572-7377}
}%

\thanks{Muhammad Umer, Muhammad Ahmed Mohsin, and Syed Ali Hassan are with the School of Electrical Engineering and Computer Science (SEECS), National University of Sciences and Technology (NUST), Pakistan.}
\thanks{Aryan Kaushik is with the Department of Computing and Mathematics, Manchester Metropolitan University, UK.}
\thanks{Qurrat-Ul-Ain Nadeem is with New York University (NYU) Abu Dhabi, UAE, and NYU Tandon School of Engineering, USA.}
\thanks{Ali Arshad Nasir is with the Department of Electrical Engineering and Center for Communication Systems and Sensing, King Fahd University of Petroleum and Minerals (KFUPM), Dhahran 31261, Saudi Arabia.}
}

\maketitle
\begin{abstract}
    Reconfigurable intelligent surface (RIS)-assisted aerial non-terrestrial networks (NTNs) offer a promising paradigm for enhancing wireless communications in the era of 6G and beyond. By integrating RIS with aerial platforms such as unmanned aerial vehicles (UAVs) and high-altitude platforms (HAPs), these networks can intelligently control signal propagation, extending coverage, improving capacity, and enhancing link reliability. This article explores the application of deep reinforcement learning (DRL) as a powerful tool for optimizing RIS-assisted aerial NTNs. We focus on hybrid proximal policy optimization (H-PPO), a robust DRL algorithm well-suited for handling the complex, hybrid action spaces inherent in these networks. Through a case study of an aerial RIS (ARIS)-aided coordinated multi-point non-orthogonal multiple access (CoMP-NOMA) network, we demonstrate how H-PPO can effectively optimize the system and maximize the sum rate while adhering to system constraints. Finally, we discuss key challenges and promising research directions for DRL-powered RIS-assisted aerial NTNs, highlighting their potential to transform next-generation wireless networks.
\end{abstract}

\section{Introduction}

Sixth-generation (6G) wireless networks promise ubiquitous and seamless connectivity, catering to the ever-growing demands of an increasingly interconnected world. With the rapid growth of data-intensive and delay-sensitive applications, such as extended reality, autonomous driving, and the Internet of Things (IoT), existing terrestrial networks face significant challenges in terms of capacity, coverage, latency, and efficiency~\cite{shen2023}. This necessitates a paradigm shift in network design towards non-terrestrial networks (NTNs), specifically aerial NTNs, which leverage a constellation of aerial platforms, including satellites and high-altitude platforms (HAPs), to augment and extend terrestrial network capabilities.

As envisioned by the Third Generation Partnership Project (3GPP) and the International Mobile Communication (IMT)-2030 framework, aerial NTNs will play a pivotal role in achieving the ambitious connectivity goals of 6G and beyond by providing resilient and sustainable communication infrastructure. Unmanned aerial vehicles (UAVs) are a key component of aerial NTNs, offering enhanced positioning freedom, cost-effective deployment and maintenance, and the ability to establish strong line-of-sight (LoS) links. UAVs can operate as aerial base stations (ABSs), aerial relays (ARs), or aerial user equipment (AUEs), each contributing to enhanced network performance through a variety of use cases~\cite{li2024},~\cite{10283826}.

Complementing the flexibility of NTNs is the transformative technology of reconfigurable intelligent surfaces (RISs). RISs are engineered surfaces comprising a large number of passive reflecting elements that can intelligently manipulate the propagation of electromagnetic waves. By dynamically controlling the phase shifts of these elements, RISs can enhance desired signals, suppress interference, and reshape the wireless channel to improve communication quality, offering promising applications in coverage extension, interference mitigation, and physical layer security enhancement~\cite{10596064},~\cite{10251110}. However, traditional, fixed terrestrial RIS (TRIS) deployments often face limitations in placement and reflection angles. Mounting RIS on aerial platforms to form aerial RIS (ARIS) overcomes these limitations. ARIS leverages the mobility of aerial vehicles to achieve dynamic positioning and their altitude to achieve panoramic full-angle reflection capabilities, thereby optimizing signal reflection and maximizing communication performance~\cite{10463697}. This flexibility allows ARIS to adapt effectively to changing channel conditions, user distributions, and environmental factors.

Optimizing the performance of ARIS requires sophisticated control mechanisms to effectively manage the complex interplay of RIS configurations, available resources, trajectories of aerial platforms, and dynamic channel conditions~\cite{9703337}. Deep reinforcement learning (DRL) emerges as a powerful tool to address these complexities. DRL algorithms can learn optimal control policies through trial and error, adapting to changing environments and maximizing long-term performance objectives. Using deep neural networks, DRL can handle high-dimensional state and action spaces, making it particularly well-suited for the intricate optimization problems inherent in NTNs\cite{10409745}. Therefore, this article explores the application and ability of DRL to enhance network performance in RIS-aided aerial NTNs.

The rest of the article is organized as follows. The following section provides an overview of RIS-assisted communications in the context of aerial NTNs. We then motivate the usage of DRL for network optimization and present a detailed description of proximal policy optimization (PPO). Next, a case study showcasing the effectiveness of hybrid PPO (H-PPO) in an ARIS-aided coordinated multi-point non-orthogonal multiple access (CoMP-NOMA) system is presented. We conclude by discussing the key challenges and emerging research directions for DRL-powered ARIS in the rapidly evolving landscape of future wireless networks.

\section{RIS-Aided Aerial NTNs: An Overview}

We commence by providing an overview of RIS-assisted communications within the context of aerial NTNs, highlighting their potential to revolutionize wireless networks by exploring their advantages, key optimization aspects, and potential use cases.

\subsection{Advantages of Aerial RIS}

ARIS presents significant advantages over TRIS due to its deployment capabilities and better performance characteristics. Unlike TRIS, which is confined to fixed locations, ARIS can be integrated with various aerial platforms, including UAVs, HAPs, and even satellites, enabling flexible and dynamic deployment. This mobility allows for on-demand coverage extension, rapid deployment in disaster scenarios, and adaptive positioning for optimal signal reflection. Moreover, the aerial nature of ARIS facilitates 360\textdegree~panoramic full-angle reflection, surpassing the typical 180\textdegree~half-space limitation of TRIS mounted on a building facade or any other vertical surface.

The elevated positioning of ARIS also leads to several substantial improvements in communication performance. Strategic aerial positioning enables ARIS to reflect signals in a way that establishes LoS links, leading to a much lower probability of signal blockage. This implies better channel quality, resulting in improved data rates and increased reliability in communication links between satellites, aerial platforms, and ground users~\cite{10559954}. The combination of deployment flexibility and superior channel characteristics makes ARIS a compelling technology for future wireless networks, particularly in scenarios requiring dynamic coverage optimization or rapid network deployment.

\subsection{An Optimization Perspective}

Optimizing the performance of RIS-assisted aerial NTNs requires a coordinated approach that considers the unique characteristics of both the RIS and aerial platforms.

\subsubsection{Passive Beamforming}

RIS utilizes passive beamforming to enhance desired signals and mitigate interference. This involves dynamically adjusting the phase shifts of the reflecting elements on the RIS to constructively combine desired signals at the receiver while suppressing undesired signals. Thus, the objective is to maximize the signal-to-interference-plus-noise ratio (SINR) and achieve higher data rates.

\subsubsection{Trajectory Control}

The trajectories of aerial platforms equipped with RIS need careful optimization to maximize coverage, minimize path loss, and avoid obstacles. Trajectory control involves determining the optimal flight paths, altitudes, and orientations of UAVs or HAPs to ensure efficient signal reflection and coverage for users. Factors such as user distribution, channel conditions, energy efficiency, and airspace regulations must be considered~\cite{10283826}.

\subsubsection{Resource Allocation}

Realizing the performance gains of ARIS-assisted aerial NTNs necessitates efficient resource allocation, which entails strategically assigning communication resources, such as power, bandwidth, and time slots, to different network entities~\cite{9703337}. Factors like user demand, channel conditions, quality of service (QoS) requirements, and energy constraints need to be considered for efficient resource utilization.

\begin{figure*}[ht!]
    \centering
    \includegraphics[width=0.9\textwidth]{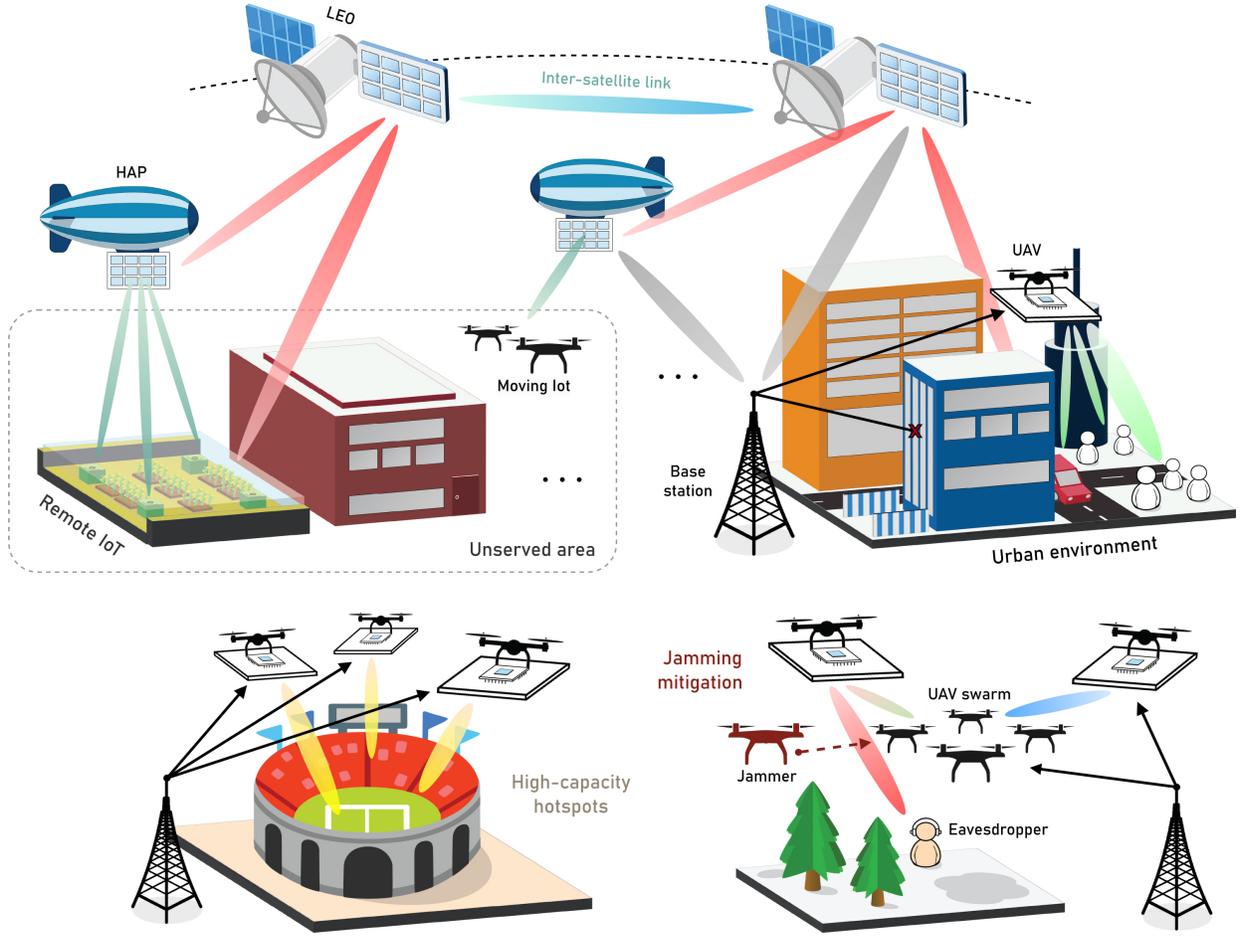}
    \caption{Illustrative use cases of RIS in aerial NTNs for enhanced connectivity, coverage, and security.}
    \label{fig:use_cases}
\end{figure*}

\subsection{Use Cases of RIS in Aerial NTNs}

The unique capabilities of RIS integrated with aerial NTNs make it suitable for various use cases, as illustrated in Fig.~\ref{fig:use_cases}.

\subsubsection{Ubiquitous Connectivity for Remote IoT}

ARIS, by leveraging the flexibility and mobility of aerial platforms like UAVs and HAPs, can extend connectivity to remote IoT devices, even in areas lacking terrestrial infrastructure~\cite{10463684}. As shown in Fig.~\ref{fig:use_cases}, a BS transmits a signal to a LEO satellite, which relays it via inter-satellite links to another RIS-equipped LEO positioned over the remote area. This LEO reflects the signal to an RIS mounted on a HAP, which directs it to the IoT devices on the ground. This ARIS and satellite communication-enabled multi-hop link provides reliable, low-latency connectivity to remote areas, supporting crucial applications like smart agriculture and environmental monitoring.

\subsubsection{Enhanced Urban Coverage}

Dense urban environments often suffer from signal blockage and multipath fading, hindering communication reliability and data rates. ARIS can be strategically deployed to overcome these challenges, as shown in Fig.~\ref{fig:use_cases}. By reflecting signals from the BS or LEO satellites, ARIS creates alternative signal paths, bypassing obstacles like tall buildings and extending coverage to shadowed areas. This dynamic positioning enables improved signal quality and extended connectivity for users in challenging urban environments.

\subsubsection{High-Capacity Hotspots}

High-density user scenarios, such as stadiums or concert venues, require high-capacity connectivity to meet the simultaneous data needs of numerous users. ARIS can effectively address these demands by intelligently reflecting and directing signals from BSs or other aerial platforms towards the high-density area. Fig.~\ref{fig:use_cases} depicts multiple ARISs hovering over a stadium, reflecting and focusing signals from a BS to provide high-quality, high-capacity connectivity to the dense user population.

\subsubsection{Secure UAV Swarms}

Secure and reliable communication is vital for UAV swarm operations, especially in sensitive applications like surveillance, data collection, and disaster response~\cite{10430396}. ARIS can significantly enhance the security of UAV swarm communications by mitigating jamming and eavesdropping attempts. Strategically positioned ARIS can create focused beams toward the intended UAV swarm while simultaneously creating nulls in the direction of potential eavesdroppers or jammers. Fig.~\ref{fig:use_cases} shows how two strategically positioned ARISs can protect a UAV swarm. One ARIS amplifies the desired signal from the BS towards the swarm, ensuring reliable communication, while the second ARIS reflects a jamming signal from a malicious drone towards an eavesdropper, thereby protecting the swarm's communication integrity.

\section{Deep Reinforcement Learning for Enhanced RIS-Assisted Aerial NTNs}

In this section, we explore the application of DRL for optimizing RIS-assisted aerial NTN communication. We motivate the use of DRL, provide a detailed explanation of a state-of-the-art DRL algorithm, and present a case study showcasing the effectiveness of the proposed DRL-based solution.

\subsection{Why DRL?}

Reinforcement learning (RL), in essence, is the science of decision making. In contrast to supervised learning, which relies on labeled data, RL involves an agent learning to make decisions through trial and error, interacting with an environment and receiving rewards or penalties based on its actions. The ultimate goal of any RL agent is to learn a policy that maximizes its cumulative reward over time. This makes RL particularly suitable for dynamic and complex systems, where it is difficult or infeasible to pre-program optimal behavior.

Deep reinforcement learning (DRL) enhances RL by incorporating deep neural networks as function approximators which are capable of handling high-dimensional state and action spaces, such as those found in complex wireless communication systems. To understand the learning behaviour of RL agents, it is essential to introduce the concepts of state value, state-action value, and policy.

\begin{itemize}
    \item \textit{State Value:} The value of a state represents the expected long-term reward the agent can achieve starting from that state and following a specific policy. It quantifies the \textit{goodness} of being in a particular state.

    \item \textit{State-Action Value:} The state-action value represents the expected long-term reward when starting in a specific state, taking a particular action, and then following the given policy.

    \item \textit{Policy:} A policy, denoted by $\pi_d$ for discrete policies and $\pi_c$ for continuous policies, is a function that maps states to actions. It dictates the behaviour of agent, guiding it to choose actions based on the observed state aiming to maximize the expected long-term reward.
\end{itemize}

\begin{figure*}[t!]
    \centering
    \includegraphics[width=0.95\textwidth]{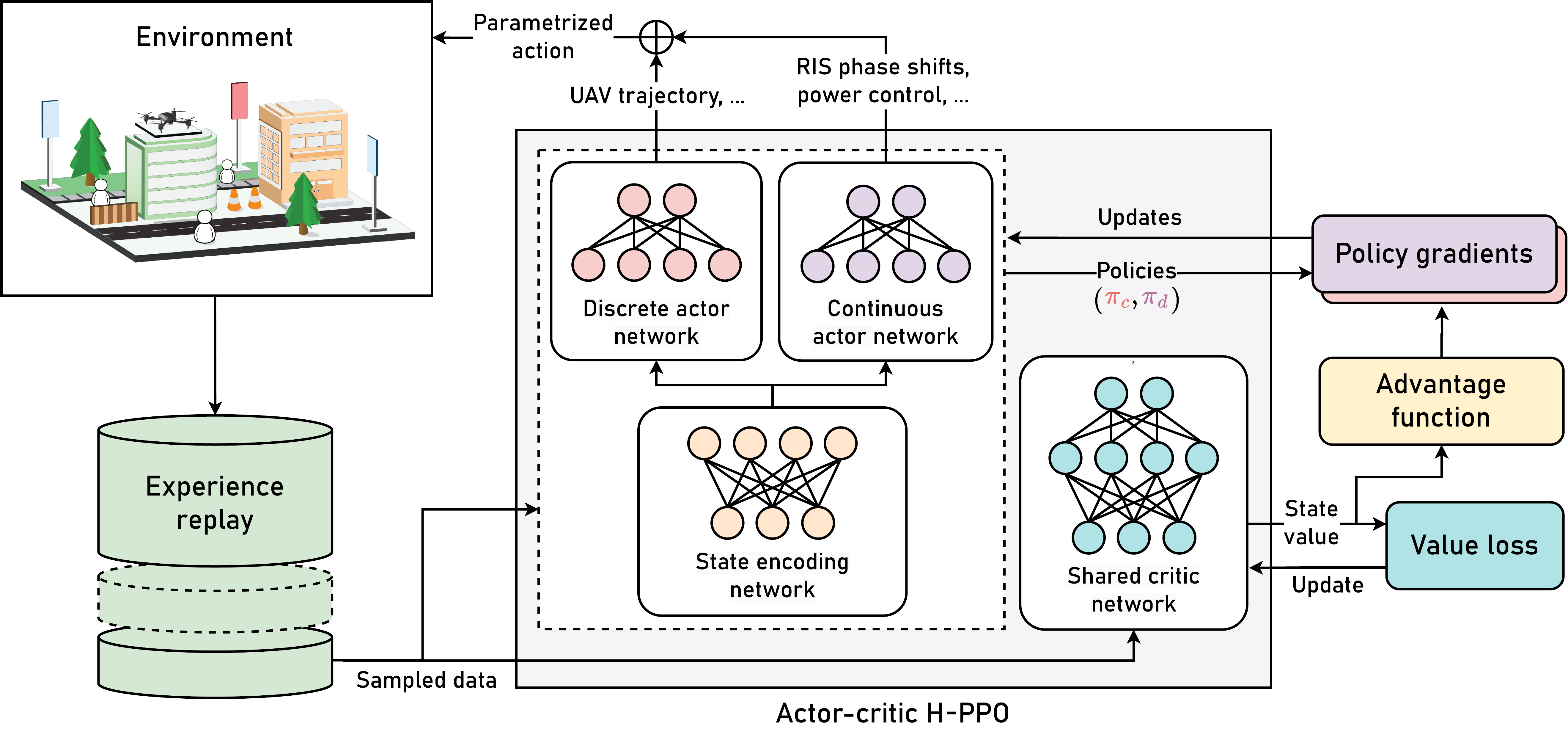}
    \caption{Architecture of H-PPO for optimization of RIS-assisted aerial NTNs.}
    \label{fig:hppo}
\end{figure*}

Optimizing RIS-assisted aerial NTNs presents unique challenges due to the inherent complexity and dynamic nature of these systems. The interplay of satellite movement, aerial platform trajectories, RIS configurations, and resource allocation strategies across high-dimensional state and action spaces demands intelligent and adaptive control mechanisms. While conventional optimization techniques, such as convex optimization, have been applied to wireless communication problems, they often struggle to cope with the dynamic and unpredictable nature of RIS-assisted aerial NTNs. These techniques typically rely on accurate and instantaneous channel state information (CSI) and involve solving computationally intensive optimization problems, leading to significant overhead and delays. This is particularly problematic in scenarios with mobile UAVs, where frequent updates to CSI and resource allocation are necessary, further amplifying the computational burden and impacting real-time performance.

DRL emerges as a powerful solution for addressing these challenges. Unlike traditional optimization approaches, DRL algorithms can efficiently learn and update control policies online, adapting to the dynamic nature of these networks where real-time adaptation is crucial. Among the various DRL algorithms, PPO stands out as a highly effective choice. As a classical policy gradient algorithm, PPO exhibits enhanced stability and adaptability, making it well-suited for navigating the challenges posed by the dynamic environments of RIS-assisted aerial NTNs. This adaptability stems from its ability to adjust the policy update step size during training, contrasting with conventional policy gradient algorithms that rely on a fixed and often challenging-to-tune step size. PPO further enhances stability and efficiency through several key mechanisms, which will be explored in detail in the following subsection. Additionally, we introduce H-PPO, an extension of PPO specifically tailored for hybrid action spaces, a common characteristic of RIS-assisted aerial NTNs.

\subsection{Description of PPO \& H-PPO}

PPO is a policy gradient-based DRL algorithm acclaimed for its simplicity, reliability, and efficiency. Unlike traditional policy gradient methods, which often suffer from instability due to large updates, PPO aims to improve the policy in a more controlled manner; policy updates in PPO are bounded by a \textit{trust region} enacted by a surrogate objective function, preventing drastic changes that could destabilize the learning process.

PPO achieves stable and efficient learning through several key mechanisms. To mitigate the high variance associated with traditional policy gradient methods, PPO employs a \textit{clipped surrogate objective function}, ensuring stability by keeping updates within a defined trust region achieved through clipping of the probability ratio between new and old policies. PPO further boosts data efficiency by performing \textit{multi-epoch updates} on each sampled batch, allowing the agent to extract more knowledge from experiences and accelerate overall learning. Furthermore, PPO makes use of the \textit{advantage function}, which estimates the relative benefit of taking a specific action compared to the average action at a given state. By prioritizing actions with higher advantages, PPO focuses on learning from the most rewarding experiences, leading to faster convergence towards an optimal policy. Lastly, PPO employs \textit{experience replay}, a common technique in DRL, where past interactions are stored in a memory buffer and randomly sampled during training. This reduces data correlation and improves learning stability by preventing the algorithm from becoming overly biased towards recent experiences.

\begin{figure*}[t!]
    \centering
    \includegraphics[width=0.85\textwidth]{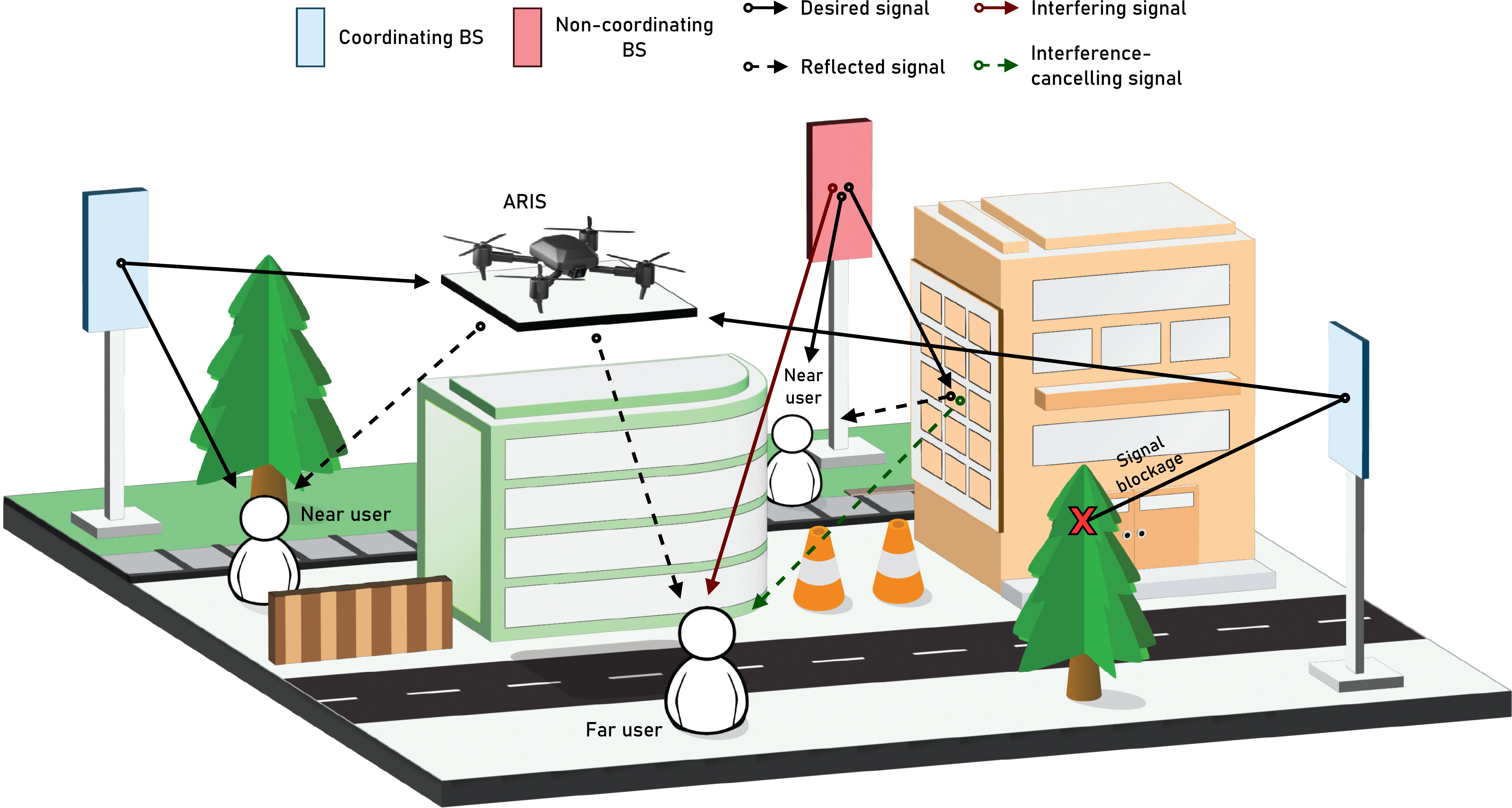}
    \caption{System model of ARIS-assisted CoMP-NOMA network.}
    \label{fig:sys_model}
\end{figure*}

Traditional RL algorithms, such as deep Q-networks (DQN) and deep deterministic policy gradient (DDPG), are primarily designed for either purely discrete or purely continuous action spaces. However, in the context of RIS-assisted aerial NTNs, we need to control both discrete actions, such as aerial maneuvers, and continuous actions, like precise adjustments of power control levels and RIS phase shifts.

Applying DQN directly to this scenario would require discretizing the continuous actions, resulting in an impractically large action space that would hinder convergence and make learning inefficient. Similarly, while DDPG can handle continuous actions, it may not be ideal for the hybrid action spaces in RIS-assisted aerial NTNs. DDPG can also exhibit instability when dealing with the complex, non-linear relationships between trajectories, RIS configurations, and communication performance, especially in dynamic environments with changing channel conditions.

To overcome this challenge, we employ H-PPO, as illustrated in Fig.~\ref{fig:hppo}. H-PPO extends the standard PPO framework by incorporating multiple output heads, allowing for simultaneous optimization of both discrete and continuous actions without resorting to excessive discretization. A shared critic network provides a common performance benchmark for both discrete and continuous actions by evaluating the value function for all states. The state encoding network processes the input state information from the sampled batch, creating a shared representation that is then fed to two separate actor heads: one dedicated to discrete actions and the other to continuous actions. While both actors interact with the same environment, their optimization occurs independently. Each actor utilizes its own distinct objective function, resulting in separate policy gradients tailored to its specific action type.

\subsection{Case Study: DRL-Enabled ARIS-Assisted CoMP-NOMA}

Building upon the evolution of cellular networks, coordinated multi-point (CoMP) techniques have been standardized to address inter-cell interference and spectrum limitations. Non-orthogonal multiple access (NOMA) further enhances spectral efficiency by allowing multiple users to share the same time-frequency resources through superposition coding of user signals at the transmitter and successive interference cancellation (SIC) at the receivers. CoMP-NOMA networks offer synergistic advantages by combining interference management through CoMP with spectral efficiency gains from NOMA. When paired with ARIS, it presents a compelling scenario for next-generation wireless networks with significant potential for enhancing coverage, capacity, and reliability.

However, such an integrated system presents a complex optimization problem due to the inherent coupling of control variables, including the trajectory of aerial platforms, RIS phase shifts, power allocation factors, and dynamic channel conditions. Conventional methods often struggle to effectively handle such coupled parameters, particularly in the context of real-time adaptation to changing wireless environments. As discussed earlier, DRL-based optimization provides an efficient and adaptable approach to managing such scenarios. The ability to learn and adapt to dynamic systems, handle high-dimensional state and action spaces, and optimize without explicit mathematical models makes it an excellent choice for addressing these challenges.

In this subsection, we demonstrate how H-PPO effectively manages the complexities of an ARIS-aided CoMP-NOMA network and how it can optimally allocate resources to maximize network sum rate.

\subsubsection{System Description}

As illustrated in Fig.~\ref{fig:sys_model}, we consider a downlink CoMP-NOMA network with both TRIS and ARIS. Three BSs are present, two of which utilize CoMP to serve a far user (FU) and are aided by the ARIS, whereas the third non-CoMP BS transmits to its own near user (NU), generating interference for the FU. We assume that direct links from the coordinating BSs to the FU are obstructed by obstacles, emphasizing the critical role of ARIS in establishing reliable communication. The signal from the non-CoMP BS to the ARIS is neglected due to the double path loss inherent in reflection links and the substantial propagation distance between them. This setup emphasizes the practical challenge of serving users in obstructed or shadowed locations while also demonstrating ARIS' ability to establish and maintain dynamic LoS links.

All communication channels in the network are modeled using the Nakagami-$m$ fading distribution with varying fading parameter $m$, providing a flexible representation of diverse channel conditions. Without loss of generality, we assume that the NUs experience better channel conditions than the FU, justifying the application of NOMA. For simplicity, we assume perfect CSI knowledge is available at the central controller, which acts as the DRL agent in this scenario. However, it should be noted that imperfect CSI due to factors such as ARIS jittering, atmospheric effects, and estimation errors is a crucial challenge for practical deployments.

The objective is to maximize the cumulative reward, defined as the network sum rate with penalties for violating operational constraints, over the UAV's operational time, which is discretized into time slots. To achieve this, the agent learns to control the UAV's trajectory, the phase shifts of both the ARIS and TRIS, and the NOMA power allocation factors. This coordinated control is subject to constraints that ensure the proper operation of SIC, maintain the UAV within a designated area of interest, and keep the phase shifts of both RISs within practical bounds.

\subsubsection{Performance Evaluation}

\begin{figure}[t!]
    \centering
    \includegraphics[width=0.95\columnwidth]{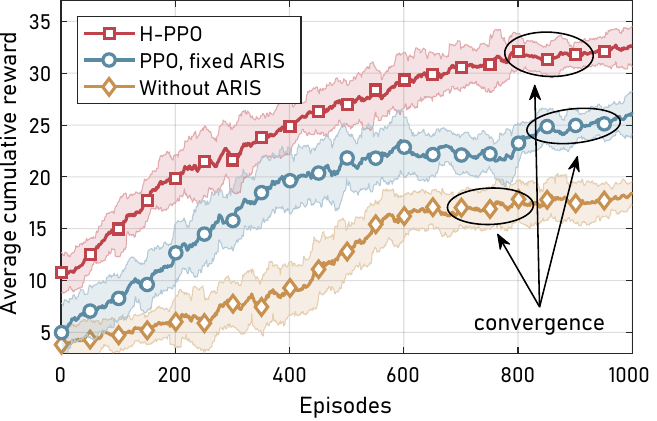}
    \caption{Average cumulative reward versus the number of training episodes.}
    \label{fig:reward}
\end{figure}

To evaluate the effectiveness of H-PPO for optimizing the ARIS-assisted CoMP-NOMA network, we conduct simulations using the following parameters unless stated otherwise. Each of the three BSs has a transmit power of $15$ dBm and serves user equipment distributed on a grid of $150$x$150$ m$^2$. The ARIS is initially positioned at the center of this grid. Additionally, the system is assumed to be operating with a bandwidth of $10$ MHz at a carrier frequency of $2.4$ GHz.

In Fig.~\ref{fig:reward}, we evaluate the average cumulative reward achieved by the DRL agent across training episodes, showcasing the convergence behavior of various DRL configurations. Both PPO and H-PPO, under different system configurations, demonstrate stable convergence, reaching a plateau in cumulative reward as training progresses. Notably, H-PPO with coordinated phase shift control of both ARIS and TRIS achieves the highest average cumulative reward. This underscores the effectiveness of jointly optimizing UAV trajectory, and passive beamforming to maximize network sum rate. Moreover, the H-PPO configuration outperforms PPO with a fixed ARIS and PPO without ARIS, highlighting the importance of dynamic ARIS positioning for establishing optimal LoS links.

\begin{figure}[t!]
    \centering
    \includegraphics[width=0.915\columnwidth]{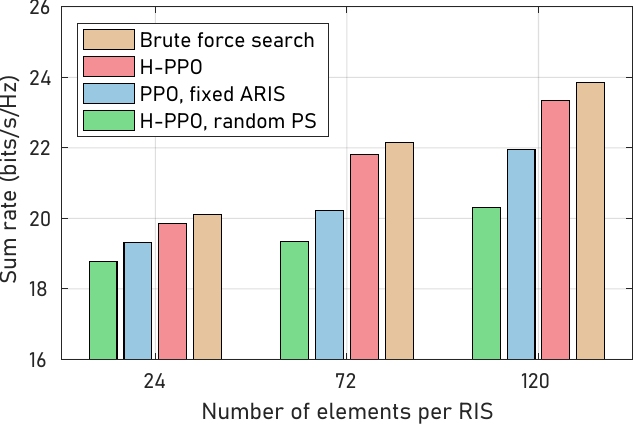}
    \caption{Network sum rate reward versus the number of elements in both ARIS and TRIS.}
    \label{fig:ele}
\end{figure}

Fig.~\ref{fig:ele} illustrates the impact of the number of reflecting elements in both ARIS and TRIS on the achievable network sum rate. To benchmark the DRL algorithms against the optimal solution, we perform a brute-force search over all possible combinations of UAV positions and RIS phase shifts, which provides a performance upper bound. As the number of elements increases, the achievable sum rate generally improves due to enhanced beamforming capabilities. As can be observed, H-PPO achieves near-optimal performance but deviates from the optimal solution as the number of elements increases due to the larger action space, which is harder for DRL agents to optimize. Comparatively, PPO with fixed ARIS and H-PPO with random ARIS phase shifts lag behind H-PPO.

\begin{figure}[t!]
    \centering
    \includegraphics[width=0.915\columnwidth]{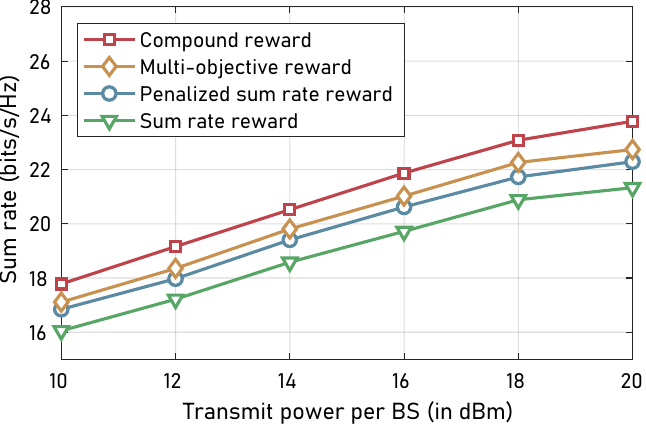}
    \caption{Network sum rate versus transmit power per BS for different reward functions.}
    \label{fig:reward_func}
\end{figure}

Fig.~\ref{fig:reward_func} shows the impact of reward function design on the network sum rate. We explored the following reward function designs.

\begin{itemize}[leftmargin=*]
    \item \textit{Sum rate reward:} Though intuitive for maximizing sum rate, this approach leads to suboptimal performance. The agent prioritizes momentarily high sum rates but may violate operational constraints, like the UAV leaving the designated area or assigning power allocation factors that hinder SIC.
    \item \textit{Penalized sum rate reward:} This design incorporates penalties for constraint violations and encourages the agent to balance sum rate maximization with constraint adherence.
    \item \textit{Multi-objective rewards:} Agent optimizes multiple objectives simultaneously, each with a distinct reward function. This allows exploration of trade-offs between conflicting goals, such as achieving high sum rates while maintaining user fairness.
    \item \textit{Compound rewards:} Unlike multi-objective rewards, this design combines multiple reward signals into a single function through weighted sums, which simplifies the learning process.
\end{itemize}

Interestingly, the highest sum rate is achieved by a compound reward function that integrates sum rate, energy efficiency, and UAV trajectory stability into a single reward signal, suggesting that holistic reward designs can be more effective.

\subsection{Complexity Analysis}

Computational complexity is of critical importance for any algorithm, as it dictates its eventual real-world deployment. For H-PPO and DRL algorithms in general, the complexity is largely determined by the size of the neural networks. The overall complexity can be approximated as $\mathcal{O}[n^2(Q_s + Q_d + Q_c)]$, where $n$ represents the number of neurons in each layer of the state encoding, discrete, and continuous actor networks, and $Q_s$, $Q_d$, and $Q_c$ denote the number of layers in each respective network. The complexity increases quadratically with the number of neurons and linearly with the number of layers. The actual computational cost depends on various factors, including hardware specifications, software implementations, and hyperparameter settings.

Mathematically analyzing the convergence of H-PPO is challenging due to the inherent complexities of neural networks and their dependence on various parameters. However, we empirically verified convergence by monitoring the average cumulative reward over numerous training episodes, as shown in Fig.~\ref{fig:reward}. Stable convergence, indicating successful learning, was observed as the reward reached a plateau. This validation provides confidence in the effectiveness of H-PPO in learning near-optimal policies in a complex environment, such as the one in our case study. It is important to note that the convergence speed drastically depends on the choice of hyperparameters, such as the learning rate, clipping parameter, and discount factor, and may require extensive tuning to achieve optimal performance.

\section{Challenges and Future Directions}

While RIS-assisted aerial NTNs hold immense potential, numerous challenges and future research directions need to be addressed to fully unlock their capabilities and enable widespread adoption in next-generation wireless networks.

\subsection{Algorithmic Challenges}

The application of DRL to optimize RIS-assisted aerial NTNs presents several algorithmic challenges that require further research. Model-free DRL algorithms like DDPG and PPO often exhibit sample inefficiency, particularly in complex scenarios like aerial NTNs where they require numerous interactions with the environment to learn effective state-to-action mappings. Improving sample efficiency through meta-learning, transfer learning, or model-based RL offers a promising direction for enabling faster training and adaptability~\cite{10525206}.

Designing effective reward functions remains another major challenge. A reward function must balance multiple, often conflicting objectives such as sum rate maximization, power minimization, user fairness, and constraint satisfaction. This typically requires many iterations of trial and error. Techniques like intrinsic motivation or curiosity-driven exploration could improve exploration strategies, especially for high-dimensional action spaces. Additionally, DRL policies trained for specific scenarios often struggle to generalize to new environments with different user distributions, channel conditions, platform configurations, or even minor changes in system parameters. Consequently, scalability to larger networks with multiple ARIS and diverse channel conditions becomes increasingly difficult. Developing agents with improved generalization capabilities and effective transfer learning mechanisms is crucial for ensuring adaptability and scalability of DRL-powered solutions in real-world deployments.

\subsection{Implementation Challenges}

Several system-level challenges hinder the implementation of ARIS in real-world scenarios. A fundamental issue lies in the accurate control of the ARIS platform and its elements, particularly due to the inherent instability of non-terrestrial networks and environmental perturbations~\cite{9771077}. Constant micro-movements and vibrations lead to phase shift errors and misalignment of RIS elements. When combined with the highly dynamic nature of aerial platforms, this introduces Doppler effects and time-varying channel conditions that complicate control mechanisms.

Imperfect CSI from estimation errors, quantization effects, and feedback delays can severely degrade overall network performance. Jittering motion of aerial platforms and atmospheric conditions further exacerbate these imperfections, introducing additional uncertainties in channel estimation. Real-world deployments must also be resilient to adversarial attacks, necessitating the development of robust beamforming algorithms and mechanisms that address security vulnerabilities in RIS control signaling and communication links. These systems must be capable of mitigating susceptibility to jamming and eavesdropping, and strong encryption techniques and sophisticated intrusion detection systems could provide the necessary safeguards.

Power consumption and energy efficiency remain major limiting factors as well. Although RIS elements themselves are passive, the overall energy expenditure of ARIS-assisted aerial NTNs is significant, especially for battery-powered UAVs. Careful planning is required for power consumption related to propulsion, control, onboard signal processing, and data transmission. Developing algorithms for optimal power allocation strategies and energy-efficient communication is essential for maximizing operational time and extending the service region of UAVs in real-world applications.

\subsection{Emerging Directions}

To advance ARIS capabilities and expand their applications, several promising research directions seem particularly compelling. Integrating RIS-assisted aerial NTNs with edge computing would bring computation and storage resources closer to users, facilitating low-latency processing, localized data management, and real-time data analytics for applications like autonomous driving, drone traffic control, and distributed AI~\cite{9771338}. This integration would support the increasing demands of data-intensive and latency-sensitive applications. Distributed learning techniques like federated learning could enable collaborative learning across multiple RIS-equipped platforms and users while protecting data privacy and enhancing scalability in large-scale deployments. This approach offers a promising way to manage the increased complexity and diverse environmental conditions encountered in vast NTN deployments by enabling collaborative learning and adaptation among distributed agents.

ARIS in integrated sensing and communication (ISAC) systems could synergistically optimize both communication and sensing functionalities and improve performance to address inherent trade-offs~\cite{10453349}. Lastly, the development of active ARIS (AARIS), which incorporates amplifiers into reflecting elements, can significantly enhance coverage, capacity, and reliability. Although active metasurfaces offer promising potential to overcome limitations of double path loss, the active components would introduce new challenges in energy and power management.

\section{Conclusion}

\label{sec:conclusion}

This article presented an overview of RIS-assisted aerial NTNs, highlighting their transformative potential in shaping the future of wireless communications. By combining the flexibility of aerial platforms with the intelligent signal manipulation capabilities of RIS, these networks offer compelling solutions to challenges in 6G and beyond. We motivated the use of DRL as a powerful tool for optimizing the complex interactions in these dynamic networks. Moreover, we presented PPO and its multi-output extension, H-PPO, as effective DRL algorithms capable of handling the hybrid action spaces often encountered in RIS-assisted aerial NTNs. A case study on an ARIS-aided CoMP-NOMA network demonstrated the superior performance of H-PPO in maximizing network sum rate.

While challenges persist in practical implementation in large-scale networks, channel estimation, robustness, and security, ongoing research actively explores solutions for implementing DRL-powered RIS-assisted aerial NTNs. Scaling DRL algorithms to manage multiple RIS-equipped aerial platforms across diverse environmental conditions necessitates further research, particularly in distributed learning and control mechanisms.



\vskip -2\baselineskip plus -1fil
\begin{IEEEbiographynophoto}{Muhammad Umer} (\href{mailto:mumer.bee20seecs@seecs.edu.pk}{mumer.bee20seecs@seecs.edu.pk}) received the B.E. degree in electrical engineering from National University of Sciences and Technology (NUST), Pakistan. His current research interests include coordinated multi-point (CoMP) transmission, non-orthogonal multiple access (NOMA), reconfigurable intelligent surface (RIS), and deep reinforcement learning (DRL).
\end{IEEEbiographynophoto}

\vskip -2\baselineskip plus -1fil
\begin{IEEEbiographynophoto}{Muhammad Ahmed Mohsin} (\href{mailto:muahmed@stanford.edu}{muahmed@stanford.edu}) received the B.E. degree in electrical engineering from National University of Sciences and Technology (NUST), Pakistan. He is currently pursuing the Ph.D. degree in electrical engineering from Stanford University, USA. His primary focus of research lies in next-generation wireless communications, deep learning (DL), and reinforcement learning (RL).
\end{IEEEbiographynophoto}

\vskip -2\baselineskip plus -1fil
\begin{IEEEbiographynophoto}{Aryan Kaushik} (\href{a.kaushik@mmu.ac.uk}{a.kaushik@mmu.ac.uk}) is an Associate Professor at Manchester Met, UK, since 2024. Previously he has been with University of Sussex, University College London, University of Edinburgh, HKUST, and held visiting appointments at Imperial College London, University of Bologna, University of Luxembourg, Athena RC, and Beihang University. He has been External PhD Examiner internationally such as at UC3M, Spain (2023). He has been an Invited Panel Member at the UK EPSRC ICT Prioritisation Panel in 2023, Editor of four books on ISAC (2024 Edition), 6G NTN (2025 Edition), ESIT (2025 Edition) all by Elsevier, Intelligent Metasurfaces (2025 Edition) by Wiley, and several journals such as IEEE OJCOMS (Best Editor Award 2024 and 2023), IEEE Communications Letters (Exemplary Editor 2024 and 2023), IEEE IoT Magazine, IEEE CTN, and several special issues. He has been invited/keynote and tutorial speaker for over 85 academic and industry events, and conferences globally such as at IEEE ICC 2024, IEEE GLOBECOM 2024 and 2023, etc., has been chairing in Organizing and Technical Program Committees of over 10 flagship IEEE conferences such as IEEE ICC 2026, 2025 and 2024, etc., and has been General Chair of over 25 workshops such as at IEEE ICC 2025 and 2024, etc.
\end{IEEEbiographynophoto}

\vskip -2\baselineskip plus -1fil
\begin{IEEEbiographynophoto}{Qurrat-Ul-Ain Nadeem} (\href{mailto:qurrat.nadeem@nyu.edu}{qurrat.nadeem@nyu.edu}) received the M.S. and Ph.D. degrees in electrical engineering from the King Abdullah University of Science and Technology, Thuwal, Saudi Arabia, in 2015 and 2018, respectively. She is currently an Assistant Professor at New York University Abu Dhabi, and NYU WIRELESS, NYU Tandon School of Engineering. Her research interests lie in the areas of wireless communications, signal processing, and electromagnetics and antenna theory.
\end{IEEEbiographynophoto}

\vskip -2\baselineskip plus -1fil
\begin{IEEEbiographynophoto}{Ali Arshad Nasir} (\href{mailto:anasir@kfupm.edu.sa}{anasir@kfupm.edu.sa}) received his the.D. degree in telecommunications engineering from the Australian National University (ANU), Australia, in 2013. He joined the Department of Electrical Engineering at King Fahd University of Petroleum and Minerals (KFUPM), Dhahran, Saudi Arabia, in 2016, where he is currently working as an Associate Professor. His research interests are in the area of signal processing in wireless communication systems. He is an Associate Editor for IEEE Communications Letters.
\end{IEEEbiographynophoto}

\vskip -2\baselineskip plus -1fil
\begin{IEEEbiographynophoto}{Syed Ali Hassan} (\href{mailto:ali.hassan@seecs.edu.pk}{ali.hassan@seecs.edu.pk}) received the Ph.D. degree in electrical engineering from the Georgia Institute of Technology (Georgia Tech), Atlanta, USA, in 2011, an M.S. in mathematics from Georgia Tech in 2011, and an M.S. in electrical engineering from the University of Stuttgart, Germany, in 2007. Currently, he is a Professor with the School of Electrical Engineering and Computer Science (SEECS), NUST, where he is the Director of the Information Processing and Transmission Lab, which focuses on various aspects of theoretical communications. He has authored more than 300 conference and journal articles, mainly in the area of wireless communications.
\end{IEEEbiographynophoto}

\end{document}